\documentclass[twocolumn,showpacs]{revtex4}

\usepackage[dvips]{graphicx}%
\usepackage{dcolumn}
\usepackage{amsmath}
\usepackage{epsfig}

\begin{document}

\preprint{gr-qc/0107092}

\title{Conventional Forces can Explain the Anomalous
Acceleration of Pioneer 10}

\author{Louis K. Scheffer}
 \email{lou@cadence.com}
\affiliation{%
Cadence Design Systems \\
555 River Oaks Parkway \\
San Jose, CA 95134
}%

\date{\today}

\begin{abstract}
Anderson, {\it et al.}\ find the measured trajectories of Pioneer 10 and 11 
spacecraft deviate from the trajectories computed from known forces acting
on them.  This unmodelled acceleration (and the less well known, but similar,
unmodelled torque) 
can be accounted for by non-isotropic radiation of spacecraft heat.  
Various forms of non-isotropic
radiation were proposed by Katz, Murphy, and Scheffer, but Anderson, 
{\it et al.}\ felt that none of these could explain
the observed effect.  This paper calculates the known effects in more detail and
considers new sources of radiation, all based on spacecraft construction.
These effects are then modelled over the duration of the experiment.
The model reproduces the
acceleration from its appearance at a heliocentric distance of 5 AU to the
last measurement at 71 AU to within 10 percent.  However, it predicts
a larger decrease in acceleration between intervals I and III 
of the Pioneer 10 observations than is observed.  This is a 2 sigma discrepancy
from the average of the three analyses (SIGMA, CHASMP, and Markwardt).
A more complex (but more speculative) model provides a somewhat better fit.
Radiation forces can also plausibly explain the previously unmodelled torques, including
the spindown of Pioneer 10 that is directly proportional to spacecraft bus heat,
and the slow but constant spin-up of Pioneer 11.
In any case, by accounting for the bulk of the acceleration, the
proposed mechanism makes it much more likely that the entire effect can be
explained without the need for new physics.

\end{abstract}

\maketitle
\setlength{\abovedisplayskip}{1mm}
\setlength{\belowdisplayskip}{1mm}
\setlength{\abovedisplayshortskip}{-0.5mm}
\setlength{\belowdisplayshortskip}{0mm}

\section{INTRODUCTION}
\label{intro}

In \cite{anderson}, Anderson {\it et al.}\ compare the measured trajectory
of spacecraft against the theoretical trajectory  computed from known
forces acting on the spacecraft.  They find a small but significant
discrepancy, referred to as the unmodelled or anomalous acceleration.
It has an approximate magnitude of $\rm 8\times 10^{-8}\;cm\;s^{-2}$ 
directed approximately towards the Sun.  Needless to say, {\it any} 
acceleration
of {\it any}  object that cannot be explained by conventional physics is
of considerable interest.  These spacecraft have been tracked very
accurately over a period of many years, so the data are quite reliable,
and the analysis, though complex, has been reproduced by Markwardt\cite{markwardt}.
Explanations for the acceleration fall into two general categories - either
new physics is needed or some conventional force has been overlooked.

One of the most likely candidates for the anomalous acceleration is
non-isotropic radiation of spacecraft heat.  This is an appealing explanation
since the spacecraft dissipates about 2000 watts total; if only 58 watts
of this total power was directed away from the sun it could account for the
acceleration.  The bulk of the spacecraft heat is radiated from the two
Radioisotope Thermoelectric Generators (RTGs), which convert the heat
of decaying plutonium to electrical power to run the spacecraft.  The
remainder of the heat is radiated from various spacecraft components
as a result of electrical power dissipation, and by a few small Radioisotope
Heater Units (RHUs) which serve to keep crucial components warm.
At least three mechanisms been proposed that could convert heat radiation
to net thrust - non-isotropic radiation from the RTGs themselves, heat 
from the RTGs reflected off the antenna, and non-isotropic radiation 
from the spacecraft bus.
Anderson {\it et al.}\ reply 
with arguments against each of the proposed mechanisms.

Although less well known, Anderson {\it et al.}\ also report that
both Pioneers experience anomalous angular accelerations.  Pioneer 10
is spinning down at a rate corresponding to a torque
of approximately $\rm 4.3\times 10^{-8}\; n-meters$ (in 1986).  
This torque is slowly decreasing - for most of the
data span, intervals I and III of Anderson, 
the torque is directly proportional to the power dissipated
by the spacecraft bus (in interval II, it appears that gas leaks
dominate the spin behavior).  This proportionality, and the size of the
effect, lead naturally to an explanation of non-symmetric radiation of
bus heat, supplemented by somewhat larger gas leaks in interval II.
Pioneer 11, when not maneuvering, was slowly and constantly spinning up. The
authors speculate that the source could be gas leaks.

This paper argues once again that non-isotropic radiation is
the most likely 
cause for both the unmodelled acceleration and at least some of the unmodelled torques.
Each of the radiation asymmetries is re-examined, and a few previously
unmodelled forces are included.  Their sum is 
more than enough to account for the acceleration, and provides a
plausible explanation for the unmodelled torques.
Furthermore, we compare the acceleration induced by the proposed mechanisms
with the measured data.  We get reasonable, but not perfect, agreement 
over the whole data span.  The main discrepancy is that the radiation
thrust is predicted to decrease more quickly than the observed acceleration.
The discrepancy is small (less than $1\sigma$) from the analysis of 
Markwardt\cite{markwardt}, but roughly a $2 \sigma$ discrepancy 
from the average results of the three analyses. 

Getting radiation forces right is notoriously difficult.  
Even for Cassini, whose construction is well known, the predicted
and measured values differ by 50\%\cite{anderson02}.
However,
the total force can be no larger than the sum of the possible components,
though it can easily be less.  Therefore the main job is to show that
enough force is available; any lesser result is easily explained.

\section{The Anomalous Acceleration}

\begin{figure}[ht]
 \begin{center}
 \noindent
 \resizebox{9cm}{!}{\includegraphics{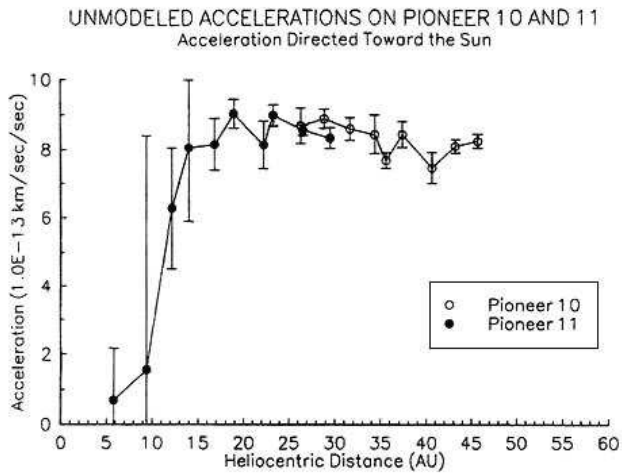}}
   \caption{Unmodelled acceleration as a function of distance from the sun, by
   Anderson {\it et al.}\ \cite{anderson02}.
    \label{fig:correlate}}
 \end{center}
\end{figure}

As the Pioneer spacecraft recede from the sun, solar forces decrease
and only gravitational forces, and an occasional maneuver, should affect the
trajectory of the spacecraft.  Anderson, {\it et al.}\ noticed that a
small additional acceleration needed to be added 
to make the measured data and computations match.  This is the anomalous
acceleration, which started to become noticeable about 5 AU from the
sun.  It was roughly the same for Pioneer 10 and 11, as shown in Figure
\ref{fig:correlate}.

\begin{figure}[ht]
 \begin{center}
 \noindent
 \resizebox{9cm}{!}{\includegraphics{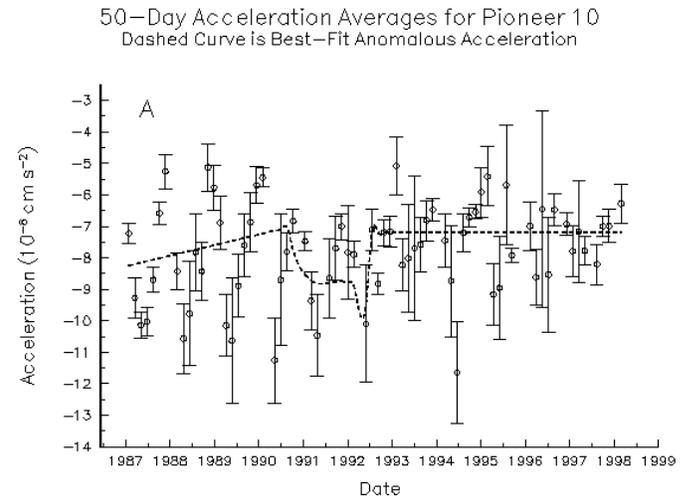}}
   \caption{Unmodelled acceleration and an empirical fit from 
   Turyshev\cite{Turyshev}.
    \label{fig:pio_turyshev}}
 \end{center}
\end{figure}

\begin{figure}[ht]
 \begin{center}
 \noindent
 \resizebox{9cm}{!}{\includegraphics{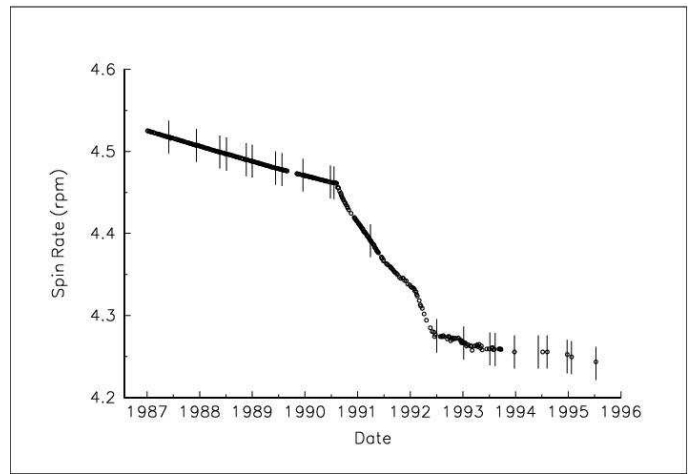}}
   \caption{Spin changes in Pioneer 10 \cite{anderson}.  The vertical
   lines indicate the times of maneuvers.
    \label{fig:spin10}}
 \end{center}
\end{figure}

Additional constraints come from further study of Pioneer 10, 
since the data are higher quality and
the data span is long enough to provide significant constraints due to
the radioactive decay of the heat sources.  Figure \ref{fig:pio_turyshev},
reproduced from \cite{Turyshev}, shows the measured acceleration 1987 to 1998.
(Although they have different horizontal axes, Figure \ref{fig:pio_turyshev}
largely follows Figure \ref{fig:correlate} chronologically.  Pioneer 10 was at
40 AU in 1987.)
The authors divide the history into three intervals. Interval I is January 1987 to
July of 1990, interval II from July of 1990 to July of 1992, and interval III is
from July of 1992 to the June of 1998.  The authors make this distinction
by looking at the spin rate of the craft (see Figure \ref{fig:spin10}).  
In intervals I and III it was decreasing
smoothly, but in interval II it decreased quickly and irregularly.  They therefore
consider the data from interval II to be less reliable than intervals I and III,
since whatever affected the spin in interval II (probably gas leaks) may also have affected
the acceleration.

More recent analyses have refined these results somewhat, though
the main conclusions remain unchanged.  Three different analyses
have been reported in the literature.
SIGMA and CHASMP are two different
trajectory modelling programs each with many possible analysis options.
We use the best Weighed Least Squares (WLS) results from each program,
from \cite{anderson02}.
Markwardt\cite{markwardt} wrote an new program with the explicit goal of an independent
re-analysis.

Table \ref{SummaryTable} shows the most recent results from \cite{anderson02},
which fits a constant, independent acceleration in each interval.
Table \ref{MarkwardtTable} shows the results of Markwardt's re-analysis
which fits a constant plus a linear term to the data from 1987-1994.  His
best solution is
$$ a(t) = -8.13\cdot 10^{-8}\mathrm{ cm/sec^2} + 3.7\cdot 10^{-17} t\; \mathrm{cm/sec^3}$$
where $t$ is the time in seconds since the beginning of 1987.
Accelerations are in units of $\rm 10^{-8}\;cm\;s^{-2}$.
For convenience, we show the amount of directed power,
in watts, that would be needed to account for each acceleration, assuming
the 241 kg estimate of spacecraft mass from \cite{anderson02}.

Note that each program claims very small formal errors, but the programs
differ from each other by far greater amounts.  Therefore the errors are
probably systematic, not random, and the differences between the programs
are better estimates of the real uncertainties.  If we create a meta-analysis
by averaging over the
3 analyses (rather dubious, but the best we can do) we get an acceleration
in interval I of $8.08\pm 0.12$ ($58.3\pm 0.87$ watts).  Agreement here
is good, so this number can
be regarded as fairly secure.
The variation with time is less clear, with SIGMA and CHASMP
showing a 2.00\% and 4.12\% decrease between interval I and interval III,
and Markwardt finding a linear trend that predicts a 10.6\% decrease in
this interval.  This gives a meta-prediction of $(5.55 \pm 3.63)$\%
decrease from interval I to III.
\begin{table}
\begin{center}
\caption{Summary of results from Anderson, {\it et al.}\cite{anderson02}}
\label{SummaryTable}
\begin{tabular}{|c||c|c||c|c|} \hline
Interval& SIGMA & equiv. & CHASMP & equiv\\
	& accel.& watts  & accel. & watts\\
\hline \hline
Jan 87- Jul 90  & 8.00 $\pm$ 0.01 & 57.8 & 8.25 $\pm$ 0.03 & 59.6 \\ \hline
Jul 92 - Jul 98 & 7.84 $\pm$ 0.01 & 56.7 & 7.91 $\pm$ 0.01 & 57.2 \\ \hline
\end{tabular}
\end{center}
\end{table}

\begin{table}
\begin{center}
\caption{Summary of results from Markwardt\cite{markwardt}}
\label{MarkwardtTable}
\begin{tabular}{|c||c|c|} \hline
date&           & equiv. \\
	& accel.& watts  \\
\hline \hline
Jan 87 - Mar 94 & 7.70 $\pm$ 0.02 & 55.7 \\
(all data, constant acc.) &                 &      \\ \hline
Jan 87 - Jul 90 & 7.98 $\pm$ 0.02 & 57.7 \\
(constant acc.) &                 &      \\ \hline
Jan 87          & 8.13 $\pm$ 0.02 & 58.8 \\
(from linear fit)&                &      \\ \hline
Jan 87 - Jul 90 & 7.93 $\pm$ 0.02 & 57.3 \\
(from linear fit)&                &      \\ \hline
Jul 92 - Jul 98 & 7.14            & 51.6 \\
(extrapolated from linear fit)  &                 &      \\ \hline
\end{tabular}
\end{center}
\end{table}

\begin{figure}[ht]
 \begin{center}
 \noindent
 \resizebox{9cm}{!}{\includegraphics{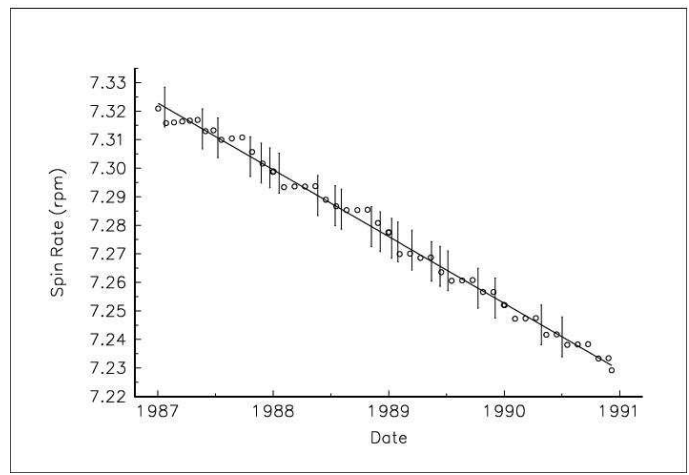}}
   \caption{Spin changes in Pioneer 11 \cite{anderson}.  The vertical
   lines indicate the times of maneuvers.
    \label{fig:spin11}}
 \end{center}
\end{figure}

Although much less publicized, there are other unmodelled forces acting on the 
craft as well.  
In the absence of external forces and/or spacecraft structure changes, 
the spin rate should not change.  
It does change, though, as shown in Figures \ref{fig:spin10} and \ref{fig:spin11}, 
from\cite{anderson02}.  Note that Pioneer 10 is spinning down
at a rate proportional to the bus power (in intervals I and III), and Pioneer
11 is spinning up, except at manuevers.
From the viewpoint of fundamental physics, unexplained torques are as 
interesting as unexplained forces, and the two are of comparable size 
(13 watt-meters and 57 watts for Pioneer 10).  Nonetheless, \cite{anderson02} assumes 
that the spin changes are caused by spacecraft systematics, but
tries to show the acceleration changes are not.  
This distinction is driven partly by the data
(the spin rates change at boundaries defined by spacecraft events, 
the acceleration does not)
but also by a lack of any remotely plausible alternative.  
There are many (and interesting) theories that could cause acceleration 
(modified gravity, dark matter, and so on) but there are few proposed theories
that could cause anomalous spins.

\section {PREVIOUS WORK}
\label{pioneer}

\begin{figure}[ht]
 \begin{center}
 \noindent
 \resizebox{9cm}{!}{\includegraphics{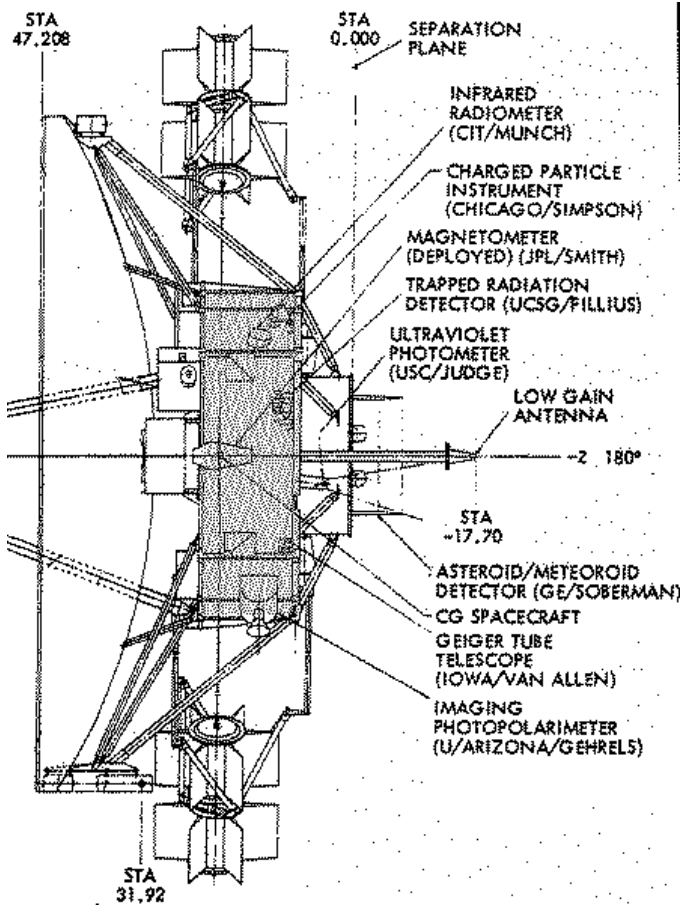}}
   \caption{Reproduction of Figure 3.1-2 from \cite{piodoc}.
   A few lines were removed for clarity, and the main equipment compartment
   is shaded in.
    \label{fig:pio_design}}
 \end{center}
\end{figure}

For the convenience of the reader, 
section \ref{sec:pio_description} consists of direct quotes from 
\cite{anderson02}, covering 
the relevant details of the Pioneer spacecraft, and 
Figure \ref{fig:pio_design}, from
\cite{piodoc}.
Many other
paper\cite{piodoc} and web\cite{pioweb,pioweb2} descriptions are available.
In section \ref{subsec:PreviousWork} 
we summarize the existing literature on the hypothesis that 
non-isotropic radiation is responsible for the unmodelled acceleration.

\subsection{General description of the Pioneer spacecraft, 
from \cite{anderson02}}
\label{sec:pio_description}

The main equipment compartment is 36 cm deep.  The hexagonal flat top and
bottom have 71 cm long sides.  
Most of the scientific instruments' electronic
units and internally-mounted sensors are in an instrument bay (``squashed''
hexagon) mounted on one side of the central hexagon.
  
At present only about 65 W of power is available to Pioneer 10
\cite{theorypower}.  Therefore,  all the instruments are no longer able to
operate simultaneously.   But the power subsystem  continues to provide
sufficient power to support the current spacecraft load: transmitter,
receiver, command and data handling, and the Geiger Tube Telescope  (GTT)
science instrument.

The sunward side of the spacecraft is the back, and the anti-sunward side,
in the
direction of motion, is the front\cite{rearfront}.  


\subsection{Gas leaks}
Gas leaks are always a prime suspect when unmodelled spacecraft accelerations are found.
As the authors themselves say ``Although this effect is largely unpredictable,
many spacecraft have experienced gas leaks producing accelerations
on the order of $10^{-7}$ cm/s$^2$''\cite{anderson02}.   Furthermore the authors think that
gas leaks are a significant part of the spin behavior.  Why then do they
think that gas leaks are not the source of the acceleration?  They present four arguments:
\begin{itemize}
\setlength{\itemsep}{-1mm}
\item The effect seems constant over long periods of time (many years).
\item The acceleration does not change as a result of thruster activity, as
many gas leaks do (as valves seat/unseat.)
\item The effect is roughly the same on two spacecraft, Pioneer 10 and 11.
\item A force big enough to cause the acceleration would cause bigger spin changes
than are observed unless it was directed along the spin axis.
\end{itemize}
In rebuttal, there are many possible sources of gas leaks, not all of which
are variable or affected by thruster activity.  
(The same authors\cite{anderson02} speculate that a gas leak 
causes the spin-up of Pioneer 11, which is also constant and unaffected by maneuvers
through 4 years.)
Furthermore, the two spacecraft were intended to be identical, so an
identical artifact such as a gas leak would not be surprising, and it might be
aligned with the axis.
In short, it would require an unusual gas leak, duplicated on each spacecraft, 
to cause the observed effect, but it is certainly allowed by physics.

In \cite{anderson02}, the error budget for gas leaks is set as follows:
First, take the biggest uncommanded spin-rate change, 
assume it was caused by gas leaks, assume the leak was at the spin thrusters,
and then increase it a little.  
Thus they are setting the budget to the biggest known
leak on this particular spacecraft.  
This is hardly a rigorous method for estimating the maximum possible
size of an unknown leak, since there could be more than one
leak, and locations other than the thrusters require a bigger 
leak for the same spin change.
Furthermore, as the authors note,  other spacecraft are known to have had larger leaks.
Clearly at least some of the authors of \cite{anderson02} are not convinced
by their own argument since they still suspect gas leaks as the cause of the
unmodelled acceleration (\cite{anderson02}, Section XII).

\subsection{Non-isotropic radiation - previous work}
\label{subsec:PreviousWork}

Murphy suggests that the  anomalous acceleration seen in the
Pioneer 10/11 spacecraft can be, ``explained, at least in part, by
non-isotropic  radiative cooling of the spacecraft.''\cite{murphy},
The main idea is that heat from the main and instrument compartments would radiate
through the cooling louvers on the front of the craft.
Anderson, {\it et al.}\ argue in reply\cite{usmurphy} that over
the data span in question the  louver doors were already closed 
(if the doors were open then the effect would surely be significant).  
They conclude ``the contribution
of the thermal radiation to the Pioneer anomalous acceleration should be 
small.''  They also argue that the spacecraft power is decreasing, but
the unmodelled acceleration is not.  Scheffer\cite{Scheffer} points out
that the front of the spacecraft has a much higher emissivity than typical
thermal blankets (even with the louvers closed), and therefore the 
majority of the heat will radiate from the front in any case.  Anderson, {\it et al.}\ 
\cite{anderson01b} dispute this, based on the emissivity data 
in \cite{piodoc}, which assigns a high emissivity to the thermal
blanket.  (This data is true but misleading - it specifies the emissivity of the
outer layer of the blanket.  This is very different from the emissivity of
the blanket as a whole, called the {\it effective} emissivity, which is quite low.
The next section has a more in-depth discussion of this point.)

Katz\cite{katz} proposes that at least part of the acceleration is
generated by radiation from the RTGs reflecting off the back of the antenna.
Anderson {\it et al.}\ in \cite{anderson99} argue that this effect 
must be small since the
antenna is end-on to the RTGs, and hence gets very little illumination.

Slusher (as credited by Anderson) proposed that the forward
and backward surfaces of the RTGs may emit non-equally.  Anderson
{\it et al.}\ 
conclude there is no credible mechanism to explain the large difference in 
surfaces that would be required if this was to explain the whole effect.
\section{Discussion}

We consider asymmetrical radiation from 5 sources - the RTGs themselves,
the two spacecraft compartments,
RTG radiation reflected from the antenna, the
Radioisotope Heater Units (RHUs) on the spacecraft, and radiation from the
feed that misses the antenna.  We also consider one modelling error, a
mis-estimation of the reflectivity of the antenna to solar radiation.

Consider thermal
radiation from the spacecraft body with the louvers closed, as they have been
since 9 AU.  An extremely simple argument shows that the electrical power
dissipated in the main spacecraft compartment must result in a significant amount of thrust.  
The Pioneer antenna points roughly at the sun, and the instrument compartment
is directly behind the antenna.  Since the antenna blocks radiation in the sunward
direction, the waste heat {\it must} be preferentially rejected anti-sunward.
Referring to Figure \ref{fig:pio_design}, a good scale model is a 60 watt
bulb about 4 cm behind a 25 cm diameter pie dish.  The dish casts a huge
shadow in the sunward direction, resulting in an average anti-sunward thrust.

However, the efficiency of conversion of heat to thrust is higher than this
simple argument indicates.
From \cite{piodoc},
``The Pioneer F/G thermal control concept consists of an insulated
equipment compartment with passively controlled heat rejection via an 
aft\footnote{What \cite{piodoc} calls aft we call the front}
mounted louver system.''  
Since even a closed louver is a much better radiator than thermal insulation,
most of the radiation occurs from the front.  It's as simple as that!

Instrument heat
may also contribute to thrust, but possibly with less efficiency. This is because
the instruments could possibly radiate at right angles to the spin
axis through their observation ports, which are not covered with
thermal blankets.  Furthermore, the science compartment is much closer to the 
edge of the dish than the main compartment, so the dish will shadow much less of any
thermal radiation generated by the science instruments.

\begin{figure}[ht]
 \begin{center}
 \noindent
 \resizebox{9cm}{!}{\includegraphics{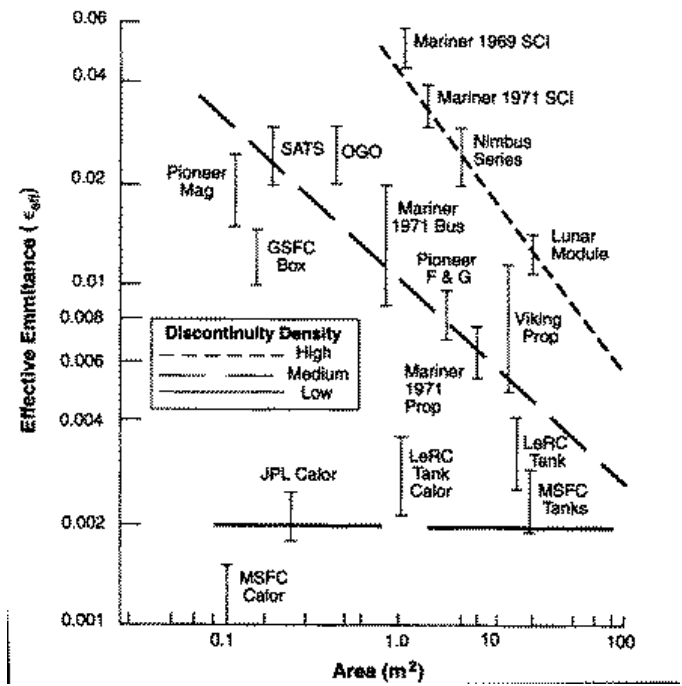}}
   \caption{Figure from \cite{Stimpson}. Pioneer F became Pioneer 10.
    \label{fig:MLI_performance}}
 \end{center}
\end{figure}

We estimate the efficiency using the spacecraft construction.
Assuming a uniform internal temperature, the power emitted from each 
surface is proportional to the area times the effective
emissivity of the surface. 
The front and back of the central equipment
compartment have about 1.3 m$^2$ area, and the sides about 1.5 m$^2$ total.
The sides and the rear of the compartment are covered with
multi-layer insulation(MLI)\cite{piodoc}.
When calculating radiation from
multi-layer insulation, the correct value to use is the ``effective'' emissivity,
$\epsilon_{eff}$,
which accounts for the lower temperature of the outer layer\cite{Stimpson}.
(Anderson\cite{anderson01b} 
points out that the
outer layer of the MLI has an emissivity of 0.70 according to \cite{piodoc}.
This is not a contradiction because 
the outer layer of the MLI is much colder than the interior - that's how MLI works.)
From \cite{Stimpson}, the multilayer insulation from 
on Pioneer 10 has an effective emissivity of 0.007 to 0.01 (see Figure
\ref{fig:MLI_performance}).
Assuming a value of 0.0085, 
and a 1998 internal temperature of 241 K\;\cite{Lasher}, the
main compartment will lose about 4 watts total through the MLI on the sides and
back (Even this may be an over-estimate.  Two of the sides are facing 1 Kw IR
sources just 2 meters away, and may even conduct heat {\it into} the compartment.).  
Allowing a few watts for conduction losses through wires and
struts, perhaps 10\% of the power (about 6 watts) goes through the
back, 10\% through the sides, and the remaining 80\% through the front.
The back radiation will have a near zero efficiency (it squirts out from
between the dish and the compartment at right angles to the flight path).
Radiation from the side should be about 10\% efficient, assuming Lambertian
radiation and a 45 degree obstruction by the dish.  Radiation from the 
front will be about 66\% efficient, again assuming Lambertian emission.
The overall efficiency of main bus radiation could therefore be as high as 54\%.

Is it reasonable for the front of the main compartment to radiate the 
47 watts or so this requires?  At an average temperature of 241 K, 
and assuming a flat surface, this
would require an average emissivity of 0.19.  From a picture of the 
Pioneer 10 replica in the National Air and Space Museum \cite{NASM}, 
the front of the spacecraft is rather complex, with considerable surface
area (such as a rather large cylinder that connects to the booster)
and a variety of surface finishes.  There are also some fairly large
instruments on the front of the spacecraft, such as the plasma analyzer
\cite{piodoc}.
Although the louver blades themselves
have a low emissivity  of 0.04\cite{piodoc}, a composite 
emissivity of 0.19 seems reasonable.

The main conclusion seems quite robust.  Multi-layer insulation is
specifically designed to reduce heat losses, whereas the louvers have at
most one layer of obstruction even when closed, and by definition are riddled
with discontinuities, which are a major source of heat leaks \cite{Stimpson}.  
The lowest emissivity material on the front, 0.04, has at least
4 times the highest quoted emissivity of the sides and back.
Surely, therefore,
a majority of the heat will be radiated from the front of the spacecraft.

\subsection{Feed pattern of the radio beam}
\label{Radio}
An ideal radio feed antenna would illuminate its dish uniformly, 
with no wasted energy
missing the dish.  However, the feed is physically small and cannot create
such a sharp edged distribution, so some radiation always spills over the edge.
Since dish area is wasted if not fully illuminated, an optimum feed (for
transmission) normally allows
about 10\% of the total power to miss the dish.  This power is converted 
to sunward thrust with an efficiency of 0.7 since it is directed
roughly 45 degree angle to the spin axis.  The rest of the energy hits
the antenna and is reflected sunward.  If $\epsilon_{FEED}$ is the fraction
of the energy that misses the antenna, and the transmitter is 8 watts, 
then the net thrust towards the sun is
$$(8~\rm w) (\epsilon_{FEED} \cdot 0.7 -(1-\epsilon_{FEED}))$$
This is negative as expected, since most of the radiation is sunward.

As a side note, the radio beam is circularly polarized and thus carries angular
momentum away from the spacecraft.  A circularly polarized beam of power $P$ and
wavelength $\lambda$ will impart a torque $T$ of
$$T = P\frac{\lambda}{hc}\hbar$$
For the Pioneer spacecraft, $P = 8$ watts and $\lambda = 13$ cm, so the torque
is about 0.165 watt-meters.  This is about a $1/100$ of the total observed
torque and can be neglected.

\subsection{Radiation from the RHUs}

From the diagrams in \cite{piodoc}, 11
1-watt (in 1972) radioisotope heater units are
mounted to external components (thrusters and the sun sensor) to keep them 
sufficiently warm.  
The diagram is not very specific, but it appears that 10 of the units, 
those mounted to thrusters and the sun sensor,
are behind the edge of the main dish.  If we assume these radiate isotropically into the
hemisphere behind the antenna, then they contribute the equivalent of
4 watts of directed force in 1998.  The remaining RHU is at the magnetometer
and would not appear to contribute net thrust.  
RHU thermal radiation will decrease with a half life of 88 years.

\subsection{Asymmetrical radiation from the RTGs}

The RTGs might contribute to the acceleration by radiating more to the front 
of the spacecraft than the rear.  This might be caused by differing solar
wind and/or dust enviroments, as proposed by Slusher.  In
\cite{anderson02} this is analyzed and estimated to be a small effect.  

However, a
more likely cause has recently been proposed by Herbert\cite{herbert}, who
notes that solar UV radiation can bleach coatings such as that used on the
RTGs, and hence make the sunward side a slightly worse radiator than the
unexposed anti-sunward side.  
There is at least some experimental evidence to indicate this is plausible.
Two commonly used coatings, Z93 and YB71, broadly similar to those on the 
Pioneer RTGs (metal oxide pigment and a silicate binder),
were tested on the NASA Long Duration Exposure Facility (LDEF) which
was left in orbit for several years and then retrieved by the space shuttle.
The IR reflectance/absorbance characteristics of the coatings were measured 
after the exposure to orbital conditions.
The Z93 coating was essentially unchanged as an IR emitter (see Figure 49 of
\cite{wilkes}), but the YB71 coating
was a worse emitter, by up to few percent depending on the wavelength 
(Figure 57 of \cite{wilkes}).  Weighting this curve by the blackbody
spectrum of a 440K radiator gives about a 3.2\% degradation in IR emission.  
This evidence is certainly not proof that this effect has occured,
since the exact coatings and environments differ (for example,
the earth orbit encounters more charged particles, atomic oxygen, and contamination,
in addition to the solar UV), but indicates that it is a plausible suspect.

In \cite{anderson02}, RTG asymmetry is found to give about 6 watts of thrust
for a 1\% asymmetry, which they take as the maximum plausible value.  
From the data above, this maximum is too low, since 3.2\% has been observed on
similar surfaces.  In this paper
we will assume that 1\% bleaching occurred, resulting in 6 watts of thrust, but
up to 3\% is certainly plausible in the absence of additional data.

Asymmetrical RTG radiation (to one side, not fore and aft) could also be 
the cause of the
slow but constant spin-up observed on Pioneer 11.  This would explain why the
spin-up is almost constant in value and unaffected by manuevers.

\subsection{Revisiting RTG reflection}

Some of the waste heat from the RTGs will reflect from the back of the high
gain antenna and be converted to thrust, as proposed by Katz\cite{katz}.
Anderson\;{\it et al.}\cite{anderson99} argue that at most 30 watts of radiation
hit the antenna, and hence RTG reflection cannot account for
the whole acceleration, which is true.  Similarly, Slabinski\cite{slabinski}, 
in an unpublished analysis from 1998, concluded that roughly 28 watts of 
radiation hits the antenna, and hence the
whole effect could not be explained.  However, it is clear the effect is real, and
can provide a significant fraction of the observed anomaly.  Only the
exact amount is in question.

The RTGs are not on-axis as viewed from the antenna.  From Figure 
\ref{fig:pio_design}, we see that
the centerline of the RTGs is behind the center of the antenna.  Measurements
from this diagram indicate this distance is about 23.8 cm. (Slabinski\cite{slabinski}
independently estimated 26 cm for this distance.)
Another figure (not included here)
from \cite{piodoc} shows the far end of the RTGs is 120.5 inches (or 3.06
meters) from the centerline.  The near end of the RTGs will then be about
60 cm further in, or at about 2.46 meters from the center.  The antenna 
extends 1.37 meters from the center, so the rim of the antenna is 69.8 cm 
off axis and 1.09 meters away radially.   Thus the edge of the antenna, where 
the illumination is by far the brightest,
views the inner RTG at an 32.6 degree angle.  This is far from on-axis.

The fins of the RTGs radiate symmetrically, and all are visible 
from the antenna, so the
center of this illumination will be 23.8 cm behind the antenna.  The
cylindrical center of the RTG is about 8.4 cm in radius \cite{Viking}
so this illumination will come from at about 15.4 cm behind the antenna.
The fins have more area than the cylinder, so for this calculation we
take a rough weighted average and
assume a cylindrical Lambertian source 20 cm behind the antenna.  We assume
the inner RTG is centered 2.66 meters from the center, and the outer RTG
2.91 meters.

\begin{figure}[ht]
 \begin{center}
 \noindent
 \resizebox{8cm}{!}{\includegraphics{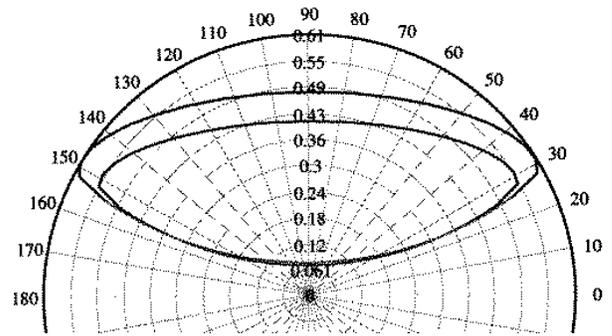}}
   \caption{Antenna size in spherical coordinates from RTGs.  Radial axis
   is angle from the centerline in radians; other axis is angle around this
   line with the magnetometer defined as zero.
    \label{fig:hit_antenna}}
 \end{center}
\end{figure}

The area blocked by the antennas is shown in Figure \ref{fig:hit_antenna} 
in spherical coordinates.  
Numerical integration of the two areas shows about 12 watts for the near
RTG and 8 watts for the far one if the total RTG power is 2000 watts.
This does not include radiation from the endcaps or supporting rods.

Combining the analyses, we conclude that at least 20 watts, but no more
than 30 watts, of radiation hits the antenna.  In this paper, we will use 25 watts as
the basis for further analysis.
This energy is turned into thrust by two effects.  First, the antenna 
shadows radiation which
would otherwise go forward.  An angle in the middle of the antenna is about
17 degrees forward; this corresponds to an efficiency of 0.3 (the true
efficiency is probably higher since the edge is both at a greater angle
and more brightly illuminated.)
Next, the energy that hits the antenna must
go somewhere.  Some will be absorbed and re-radiated; some will bounce into 
space, and some
will bounce and hit the instrument compartment, and be reflected or re-radiated
from there.  A detailed accounting seems
difficult, but an overall efficiency of 0.6 to 0.9 seems reasonable (0.3 for shadowing
and 0.3 to 0.6 for reflection and re-emission).

\subsection{Total of all effects}

Here we sum the maximum value of  all the effects as of 1998.  
The total is more than enough to
account for the acceleration, giving us the freedom to reduce some
of the efficiencies if needed to fit the data.

\begin{table}[ht]
\begin{center}
\caption{Available thrust from different sources as of 1998}
\label{ThrustTable}
\begin{tabular}{|c|c|c|c|c|} \hline
Source of effect & Total Power & Effic. & Thrust & Decay\\
\hline \hline
Rad from RHUs    & 8           & 0.5       & 4   & 0.78\%/year \\ \hline
Antenna shadow   & 25          & 0.3       & 7.5 & 0.68\%/year \\ \hline
Antenna radiate  & 25          & 0.6       & 15  & 0.68\%/year \\ \hline
RTG asymm.       & 2000        & 0.009     & 18  & 0.68\%/year \\ \hline
Feed pattern     & 0.8         & 0.7       & 0.6 & 0\% \\ \hline
Radio beam       & 7.2         & -1        & -7.2& 0\% \\ \hline
Rad., main bus   & 59          & 0.54      & 32  & see text  \\ \hline
Rad., instr.     & 1           & 0.1       & 0.1 & see text \\ \hline
Total            &             &           & 70.0  &     \\ \hline
\end{tabular}
\end{center}
\end{table}

\subsection{Antenna solar reflectivity}

In this section we argue that a mismodelled solar reflection might account
for the sudden onset of the anomalous force shown in figure 1.
This argument is offered {\it only} as a possible explanation of the
{\it onset} and initial decrease of the anomalous acceleration; it is not relevant to the
existence, magnitude, or source of the acceleration at later times since past about 30 AU
the contribution from the solar radiation is negligable.

First, we show there is surely a possibility of error in these coefficients since
the numbers for the two spacecraft disagree.
We start with the data from Anderson, averaging the SIGMA and CHASMP values.
We assume, following Anderson (section VII-B), that the trajectory was fit correctly but the
mass used in the calculation was incorrect.
We correct the fitted values using the best available estimates for
spacecraft mass, keeping the acceleration (and hence trajectory) the same.
\begin{table}[ht]
\begin{center}
\caption{Solar reflectivity from Anderson, {\it et al.}\cite{anderson02}}
\label{SolarTable}
\begin{tabular}{|c||c|c||c|c|} \hline
Spacecraft & fitted $\mathcal{K}$  & mass used & actual    & resulting \\
	   &                       & in fit    & mass      & true $\mathcal{K}$\\
\hline \hline
Pioneer 10 & 1.73                  & 251.8     & 241       & 1.66 \\ \hline
Pioneer 11 & 1.83                  & 239.7     & 232       & 1.77 \\ \hline
\end{tabular}
\end{center}
\end{table}
We would expect an nearly identical value of $\mathcal{K}$ 
for both spacecraft - they were the same
size and painted with the same paint - but we observe $\mathcal{K}=1.66$ for one and
$\mathcal{K}=1.77$ for the other, a 6.6\% difference.  
One possible explanation is that the two
spacecraft had different amounts of thermal radiation thrust, and the fitting
procedure used an adjusted value of $\mathcal{K}$ to fit the observed trajectory.

Is it possible that the trajectory is right, but $\mathcal{K}$ is wrong?
Anderson {\it et al.}\cite{anderson01b} claim
that any appreciable error in this value would have resulted in navigation errors,
but the 6.6\% difference between the Pioneers (surely not physically present) 
was easily absorbed into the fit (perhaps into the velocity increments at maneuvers, 
for example).
This would certainly not be the first time that
an excellent fit to the data was obtained with the wrong explanation.

The analysis in Anderson\cite{anderson02} reinforces this point.  They tried to
separate a constant anomalous force from the $1/r^2$ solar reflectivity, 
but found they were tightly correlated. 
On Ulysses, for example, the correlation between these two parameters was
0.888, so 90\% of the change in one parameter could be explained by a spurious
change in the other.  If the hypothesis of this article is correct, the two
parameters will be even harder to separate, since for Pioneer the
both the radiation contribution and the solar reflection are decreasing as the
spacecraft recedes from the sun (the total power is
decreasing, and the efficiency will decrease as well as the louvers close).

In the scenario of this paper, the acceleration has existed all along, and might even have
been stronger closer to the sun.  When Pioneer was closer to the sun, though,
the fitting programs absorbed the extra acceleration by adjusting the
value of $\mathcal{K}$ and perhaps other
parameters such as the delta-v of maneuvers.
As Pioneer receded from the sun, and maneuvers became less frequent, 
adjustments to these parameters could no longer fit the trajectory properly.
At this point the the anomalous acceleration ``appears''.  This argument
is not specific to radiation induced acceleration - any small radial
acceleration can be compensated for by adjusting the value of $\mathcal{K}$.

In this paper, we model the effect of any error in $\mathcal{K}$ 
by introducing a fictitious
force, whose value is simply the solar force on the
spacecraft times the error in $\mathcal{K}$.

\section{Comparison with experiment}

How well does this explanation account for the acceleration?  The
explanation has 6 adjustable parameters.
\begin{itemize}
\setlength{\itemsep}{-1mm}
\item
$\epsilon_{RHU}$, the fraction of RHU heat converted to thrust
\item
$\epsilon_{RTG}$, the fraction of RTG heat converted to thrust.  Includes
both direct asymmetry and reflection from the antenna.
\item
$\epsilon_{FEED}$, the fraction of RF power that misses the antenna
\item
$\epsilon_{INST}$, the fraction of instrument heat that is converted to thrust
\item
$\epsilon_{BUS}$, the fraction of main compartment heat that is converted to thrust
\item
$K_{SOLAR}$, the amount by which the solar reflection constant is underestimated.
This cannot exceed about 0.2 since the true value can be no more than 2.0\;.
\end{itemize} 
In theory all parameters are separable
since they decay at different rates.  In practice there are many similar
solutions that cannot be distinguished by the existing data.

We compute the net thrust as follows: let $d$ be the date in years.  
The total electrical power, in watts, is modelled as 
$$E(d) = 68+2.6\cdot (1998.5-d)$$
The RHU power, in watts, is $$ RHU(d) = 10.0\cdot 2^{-(d-1972)/88}$$  
The RTG heat dissipation, in watts, is $$RTG(d) = 2580\cdot 2^{-(d-1972)/88}-E(d)$$
We assume the distance from the sun, measured in AU, increases linearly from 20
AU in 1980 to 78.5 AU in 2001:
$$ r(d) = 20 + (d-1980)/21\cdot (78.5-20) $$
The power incident upon the antenna, in watts, is
$$ SOLAR(d) = \pi (1.37 ~{\rm m})^2 f_\odot/{r^2(d)} $$
where $f_\odot=1367 ~{\rm W/m}^{2}$(AU)$^2$
is the ``solar radiation constant'' at 1 AU.  
We use the expression from section \ref{Radio} for the radio thrust.

The power dissipated in the instrument compartment, $INST(d)$, is given 
in Table \ref{InstPowerTable} from \cite{nssdc}. 

\begin{table}[ht]
\begin{center}
\caption{Instrument power 1987-2001.  IPP = Imaging Photopolarimeter,
TRD = Trapped Radiation Detector, PA = Plasma Analyzer}
\label{InstPowerTable}
\begin{tabular}{|c|c|c|} \hline
Dates & Watts & Notes \\
\hline \hline
Jan 87 - Oct 93 & 11.6 & IPP off Oct 93\\ \hline
Oct 93 - Nov 93 & 8.1  & TRD off Nov 93\\ \hline
Nov 93 - Sep 95 & 5.3  & PA off Sep 95 \\ \hline
Sep 95 - present& 0.8 & Only Geiger active \\ \hline
\end{tabular}
\end{center}
\end{table}

The other units that were turned off during this period (the Program Storage 
and Execution unit, and the Duration and Steering Logic) 
did not affect the instrument heat since they simply substituted one heat 
source in the main compartment for another.

The electrical power that does not go into the instruments or the 
radio beam goes into the main compartment:
$$BUS(d) = E(d)-INST(d) - 8.0$$

We sum the individual sources, then convert to acceleration by dividing by 
$c$, the speed of light, and $m$, the spacecraft mass (here 241 kg):
\begin{eqnarray*}
\lefteqn{acc(d) = \frac{1}{c\cdot m} [\epsilon_{RHU}\cdot RHU(d) + } \\
 & & \epsilon_{RTG}\cdot RTG(d) + \\
 & & (8~\rm w) (\epsilon_{FEED} \cdot 0.7 -(1-\epsilon_{FEED})) + \\
 & & \epsilon_{INST} \cdot INST(d) +  \\
 & & \epsilon_{BUS} \cdot BUS(d)  - \\
 & & K_{SOLAR}*SOLAR(d)]
\end{eqnarray*}

To examine the fit, we use the plots from  \cite{anderson02,Turyshev}, and 
try to fit them with our model.  We make three fits.  
The first is a conservative fit, using only known and documented spacecraft
characteristics.  The second is the nominal fit, adding in effects such as
RTG asymmetry that are plausible but not proven.  The third
is constructed to
get the best possible fit to the data, but might be physically unrealistic.

The conservative fit uses only {\it known} and {\it documented} spacecraft
characteristics.  These are that the front of the spacecraft is a better
radiator than the sides\cite{Stimpson,piodoc}, and that the antenna will block and reflect
some of the RTG radiation\cite{piodoc,anderson02,katz,slabinski}.  
A good fit is obtained with
\begin{itemize}
\setlength{\itemsep}{-1mm}
\item
$\epsilon_{RTG} = 0.01$.  25 watts hit the antenna, 30\% blockage
efficiency, and 50\% reflection efficiency.
\item
$\epsilon_{INST} = 0.51$.  Instruments same as main bus for simplicity.
\item
$\epsilon_{BUS} = 0.51$.   About 80\% the main bus heat goes out the front, 
with Lambertian efficiency.
\end{itemize}
This model correctly predicts 58.6 watts in interval I, but predicts a decrease
in interval III to 48.6 watts.  This is a 17\% decrease as opposed to the
3\% measured in Anderson and 10.6\% of Markwardt.  
This model does not explain the onset at 5 AU, and overpredicts the rate of
decrease, but it shows that at most 20\% of the effect can be due to new physics.
At the very least, 80\% of the effect can be accounted for by entirely conventional
physics, based on known, documented, and measured spacecraft construction.

The nominal fit adds radiation from the RHUs, asymmetrical radiation from the
RTGs, feed spillover, and solar reflectance mis-modelling.  These sources are
all plausible but neither proven or disproven by any records or measurements
found so far.  The fit assigns the same efficiency to main compartment
heat and instrument heat.  This avoids much of the need to look at spacecraft
construction details and instrument history, since the acceleration only
depends on the total electrical power.  

The additional sources allow a better fit since
RHU and RTG heat decays more slowly than electrical heat, feed spillover
does not decay at all, and we can now model the onset of the anomalous acceleration
at 5 AU.  Once again, many parameter choices give similar results.
We get a reasonable fit over the entire data span with the following coefficients:
\begin{itemize}
\setlength{\itemsep}{-1mm}
\item
$\epsilon_{RHU} = 0.5$, the RHUs radiate like point sources behind the antenna.
\item
$\epsilon_{RTG} = 0.016$.  0.3\% RTG asymmetry, 30\% blockage
efficiency, and 50\% reflection efficiency.
\item
$\epsilon_{FEED} = 0.1$.  10\% of the feed power misses the antenna
\item
$\epsilon_{INST} = 0.39$.  Instrument heat radiates as main bus heat for simplicity.
\item
$\epsilon_{BUS} = 0.39$.   About 60\% of the main bus heat goes out the front, 
with Lambertian efficiency.
\item
$K_{SOLAR} = 0.2$.  Antenna reflection estimates are too low by 0.2.
\end{itemize} 

\begin{figure}
 \begin{center}
 \noindent
 \resizebox{9cm}{!}{
  \includegraphics[trim=1.2cm 0.4cm 0cm 0cm, clip=true]{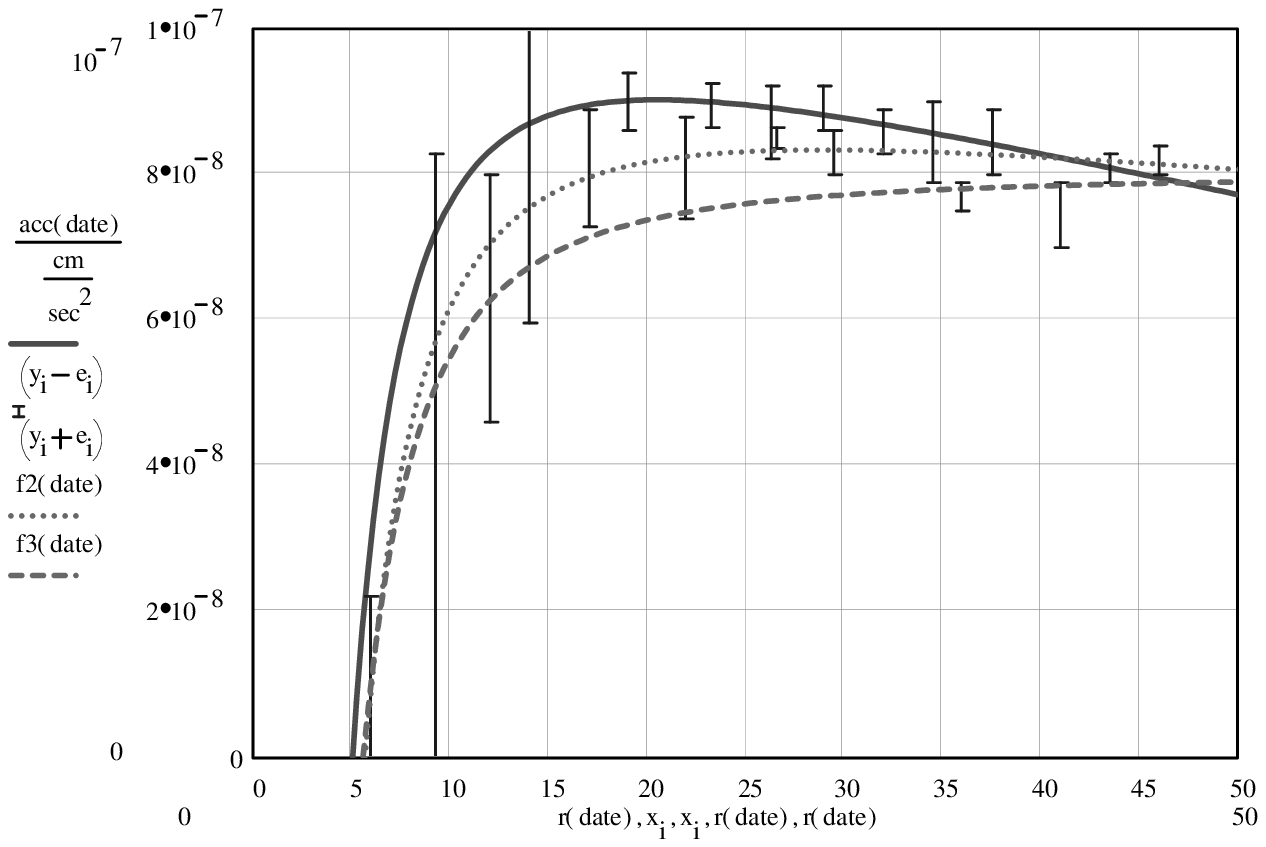}}
   \caption{Data from Figure \ref{fig:correlate} (error bars),  model prediction
   from this paper (solid line), pure 88 year half life plus solar constant error (middle line),
   and constant acceleration plus solar constant error(lowest line)
    \label{fig:EarlyAnom}}
 \end{center}
\end{figure}

\begin{figure}
 \begin{center}
 \noindent
 \resizebox{9cm}{!}{\includegraphics{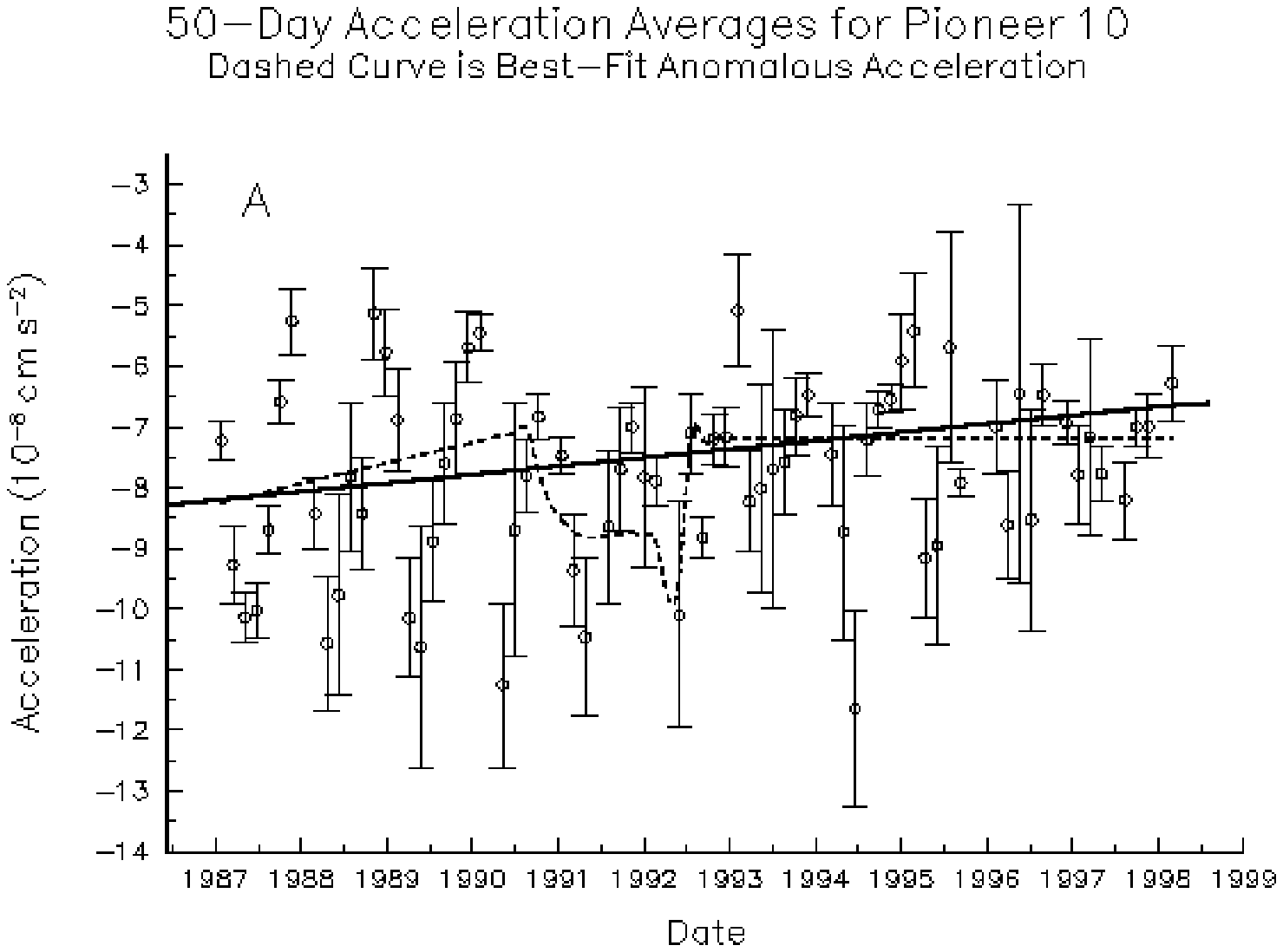}}
   \caption{Figure from \cite{Turyshev}, with fitted data added.
   The dotted line is Turyshev's empirical fit; the solid line is the
   model hypothesized in this paper.
    \label{fig:theory_vs_practice}}
 \end{center}
\end{figure}

The fit to the data is shown in Figures \ref{fig:EarlyAnom} and 
\ref{fig:theory_vs_practice}.  
The agreement seems reasonable in both regimes.  
In particular, the early 
anomalous acceleration  between 15 and 40 AU is fit well by this model.  In Figure
\ref{fig:EarlyAnom} two other models are shown, all assuming that a $1/r^2$ error
of some sort (here solar constant mismodelling) is responsible for the onset.  
The middle trace assumes the acceleration is a pure
exponential with an 88 year half life.  This is the form
for a model that assumes RTG radiation (direct or reflected) is asymmetric 
but spacecraft electrical heat is radiated isotropically.  Between 15 and 40 AU this model
underpredicts the observed decrease, where the nominal model fits much better.
This strongly favors a model where radiation from the spacecraft bus is a major
contributor to the anomalous acceleration.
The lower trace is a constant acceleration
plus an error that scales as $1/r^2$.  This shows that if the acceleration
is indeed constant at large distances, a different explanation for the onset
is required.

The fit from 1987 to 1998, shown in Figure \ref{fig:theory_vs_practice}, 
also looks reasonable.
We compare this model to the consensus of the 
most recent analyses\cite{anderson02}\cite{markwardt}.
Using the parameters above,
the average sunward thrust is 58.0 watts in interval I, and 50.2 watts in
interval III.  We can adjust the the parameters to get the correct overall
average, or the right acceleration in interval I, but in either
case we would expect to see a 13.2\% decrease from
interval I to III, where only a 5.6\% decrease is observed.
The 7.6\% discrepancy is about
2 standard deviations out.  Taken at face value, this makes it unlikely 
at about the 2\% level that this hypothesis alone accounts for all the measured result.
However, the re-analysis by Markwardt\cite{markwardt} 
has concluded that the data does not rule out a slowly decreasing
force, at least if the the decrease is an exponential with 
a half life of more than 50 years.  (This corresponds to a 9\% decrease
in the 6.75 year span between the midpoints of intervals I and III).
The decrease here is not strictly
exponential, but is close in in size and shape to that
explicitly allowed by Markwardt.

More speculatively, an even better fit to the acceleration data can be obtained by assigning different
efficiencies to instrument heat and main compartment heat.  For example,
\begin{itemize}
\setlength{\itemsep}{-1mm}
\item
$\epsilon_{RHU} = 0.5$, the RHUs radiate like point sources behind the antenna.
\item
$\epsilon_{RTG} = 0.01425$.  0.3\% RTG asymmetry, 30\% blockage
efficiency, and 60\% reflection efficiency.
\item
$\epsilon_{FEED} = 0.1$.  10\% of the feed power misses the antenna
\item
$\epsilon_{INST} = 0.40$.  About half the main bus heat radiated forwards
with Lambertian efficiency.
\item
$\epsilon_{BUS} = 0.10$.   Instruments radiate mostly to the side
\item
$K_{SOLAR} = 0.2$.  Antenna reflection estimates are too low by 0.2\;.
\end{itemize} 
gives a better fit, with only a 4.9\% discrepancy (10.5\% predicted versus 5.6\% measured) 
on the I-III decline and a
roughly equivalent fit at earlier times.  This is only about 1.3 standard deviations
from the consensus model, and an almost perfect fit for Markwardt.
However, figuring the maximal reasonable
difference between instrument efficiency and and main compartment efficiency
is difficult\cite{anderson01b}.  
One the one hand the two compartments are separate, the
instrument bay is closer to the edge of the antenna, and it has has 
side facing ports that extend though the thermal blankets.
On the other hand the two compartments are
radiatively and conductively coupled.
Without a much more detailed analysis it's very hard to determine the
maximum plausible difference in efficiencies.

Finally, asymmetric radiation offers a parsimonious explanation for both the
anomalous acceleration and the anomalous torque.  Anderson\;{\it et al.}\ note that
the Pioneer 10 spin-down torque is almost perfectly correlated with the main bus power.  
Radiation from the front of the craft, as proposed here, explains this.
The needed emission geometry is numerically
plausible - in 1986, there were 97 watts available, and 13 w-meters of torque
measured.  Assuming the radiation is emitted 50 cm from the axis (the louver location), 
if the radiation was
canted at an average angle of 15.5 degrees from the normal to the
surface, it could provide the observed torque.  Such an angle would decrease the
conversion of power into thrust by only 4\%, leaving that argument intact.
The louvers, covering the front surface and all canted to one side when closed,
provide a natural explanation for the asymmetry required.  One obvious objection
to this explanation is that it predicts Pioneer 11 should be spinning down as
well, instead of the spin-up that is actually observed.  This is not a serious
problem since the unknown spin-up mechanism, possibly gas leaks or RTG asymmetry, can easily
overpower the small torque induced by main bus radiation.

In any case, the proposed explanation, by accounting for the bulk of the effect,
makes it more likely that conventional physics can account for the entire 
unmodelled acceleration.
Conventional explanations for the remaining discrepancy
include other unmodelled effects such as
gas leaks, inaccuracies in the simple thermal model, or
the effects of
a complex fitting procedure applied to noisy data.
   
\section{Conclusions and future works}

No new physics is needed to explain the behavior of the Pioneer spacecraft.
Either gas leaks or thermal radiation, or a combination of the two, 
could explain both the linear and angular accelerations that are measured.

A strong thermal effect is certainly present, based only on the construction of the
Pioneers.  Estimates show
it can account for the magnitude of the unmodelled acceleration to within 
the errors, but overpredicts the rate of change.
The antenna shadowing of main compartment radiation
and the radiation from the RTGs falling on the antenna seem particularly
robust sources of acceleration since they are only based on geometry.  
These effects alone account for
more than half the acceleration.  The other sources - RHU radiation, differential
RTG radiation, and differential emissivity - depend more on construction
details, but all seem plausible.

This explanation also explains some other puzzles: the values of acceleration
of Pioneer 10 and 11 would be expected to be similar, but not identical,
as observed. The acceleration and the observed torque (on Pioneer 10)
share a common origin,
and the torque is proportional to the main bus heat, as observed.
Other spacecraft, built along the same general principles,
would be expected to show a similar effect, but planets and other large
bodies would not, as is observed.

The hypothesis here predicts an eventual, unambiguous decrease in 
the anomalous acceleration.
If the acceleration remains constant, on the other hand, the
hypothesis will be refuted.  
Extending the
analysis of Markwardt to the whole Pioneer data span would be useful, since
it it currently stops at 1994 and it directly includes the possibility 
of a non-constant acceleration.  Extending the analysis of Anderson by
including post 1998 data would be helpful as well.

If Pioneer 10 remains operational, additional data may allow us to improve
our understanding of the unmodelled acceleration.  The difference between
constant acceleration and the decrease predicted by the hypothesis of this paper
grows quadratically with time.  Since the beginning of data in 1987, by
2002 the two solutions differ by 4.4 cm/sec, or a doppler shift of 0.58 Hz.
Thus a single good 2002 measurement could tell the two hypotheses apart.
Unfortunately the signal is now very weak, to the point where the
standard JPL receivers have trouble locking onto the signal\cite{pioweb2}.
Careful recording of the return signal
might probably work, though, with the frequency recovered through long 
averaging.  Bigger telescopes such as Arecibo, the VLA, or Greenbank, might
conceivably be pressed into service as well.

More detailed modeling, using the Pioneer materials, construction details,
and history, could provide a much better estimate of the magnitude of this
effect.  A suitably detailed thermal model, measured in a cold vacuum
chamber, would provide the strongest evidence for or against this hypothesis.

Longer term,
other proposed experiments such as LISA\cite{LISA} are designed specifically
to reduce the systematics that bedevil retrospective analyses like Pioneer.  
(LISA is expected to be about $10^5$ times better in this respect.)  If
the anomalous acceleration is not detected in these more precise experiments, 
then almost surely the unmodelled acceleration of Pioneer 10
is caused by overlooked prosaic sources such as those proposed here.

\section{Acknowledgements}

I'd like to thank Edward Murphy and Jonathan Katz for comments and 
suggestions on an earlier version of this document; 
Edward Murphy also sent copies of the documents he found while 
investigating the same effect.  Larry Lasher and Dave Lozier of the Pioneer
project were kind enough to answer questions about the probe.  John Anderson
suggested adding the statistical likelihood calculations.  Victor Slabinski
forwarded his own independent antenna reflection calculations, and George
Herbert proposed the RTG bleaching hypothesis.  An anonymous referee
provided useful suggestions.

%



\begin{thebibliography}{99}
\bibitem{theorypower}
This is a ``theoretical value,'' which does not account for inverter
losses, line losses, and such. It is interesting to note
that at mission acceptance, the total ``theoretical'' power was 175 Watts.
\bibitem{Lasher} This temperature is the average of the 4 thermistors in
the main compartment, from telemetry on 29 Sept 98.  This data was
provided by Larry Lasher of the Pioneer project.
\end{thebibliography}
\end{document}